\documentclass[times,authoryear]{elsarticle}

\usepackage{jasr}
\usepackage{framed,multirow}

\usepackage{amssymb}
\usepackage{latexsym}
\usepackage{booktabs}
\usepackage{amsmath}
\usepackage{subfig}

\usepackage[switch]{lineno}

\usepackage{url}
\usepackage{xcolor}
\definecolor{newcolor}{rgb}{.8,.349,.1}

\usepackage[citebordercolor=white]{hyperref}

\journal{Advances in Space Research}

\begin{document}

\verso{Barman \textit{et. al}}

\begin{frontmatter}

\title{Identification of recurrent novae from parametric modeling of the optical light curve\tnoteref{tnote1}}%
\tnotetext[tnote1]{}

\author[1]{Rajarshi \snm{Barman}\corref{cor1}}
\ead{rajarshiwork123@gmail.com}
\author[1]{Nirupam \snm{Roy}\fnref{fn1}}
\ead{nroy@iisc.ac.in}
\fntext[fn1]{}

\affiliation[1]{organization={Indian Institute of Science},
                addressline={C.V. Raman Road, Mathikhere},
                city={Bangalore},
                postcode={560012},
                country={India}}

\received{}
\finalform{}
\accepted{}
\availableonline{}
\communicated{}

\begin{abstract}
Novae, characterized by sudden brightening in binary star systems, are categorized into classical novae (CNe) and recurrent novae (RNe) based on their recurrence timescales. However, identifying RNe, which occur within 100 years, presents observational challenges. A reasonable signature of RNe has been theorized and statistically validated to address this—the presence of a plateau in the optical light curve. Among the  known RNe, except T CrB and V3890 Sgr displaying S-class light curve, the rest 9 out of 11 have a P-class light curve. But classical novae can also present plateaus, which further complicates the problem of distinguishing CNe from RNe just based on plateau. Hence, in this study, we aim to conduct a phenomenological analysis of P-class light curves to comment on the recurrence nature of novae. We utilize data primarily from the AAVSO database and identify a parameter space to represent all P-class light curves, anticipating distinct parameter distributions for CNe and RNe. Analysis of parameter distributions successfully distinguishes RNe from CNe and reveals potential connections with white dwarf mass and mass accretion rate, which are the key factors. Our method indicates KT Eri to be a recurrent nova, consistent with a recent study, despite only one observed outburst. The analysis also indicates the possibility of recurrence for V2860 Ori, a prediction that may be tested by deep search for nova super-remnant for this source. Our method demonstrates the feasibility of distinguishing P-class classical and recurrent novae based solely on the optical light curve. As observations of new novae increase, this method holds promise for more precise predictions of nova recurrence nature in the future.
\end{abstract}


\end{frontmatter}


\section{Introduction}
\label{sec1}
A nova outburst occurs within a binary system where a white dwarf accretes hydrogen-rich material from its companion star, which could be a main-sequence, sub-giant, or red-giant star. The accreted material accumulates on the white dwarf's surface, forming an envelope. The white dwarf's intense surface gravity induces partial degeneracy in the accumulated material. As a result, compressional heating causes the temperature at the base of the envelope to rise. Due to degeneracy, the material cannot expand in response to the temperature increase, leading to a continuous temperature rise. When the temperature reaches approximately $10^6 - 10^7$ K, nuclear fusion ignites, causing an exponential increase in temperature. The temperature reaches the Fermi temperature ($\sim 70$ million K) very rapidly, lifting the degeneracy and resulting in a rapid change in the equation of state. This sudden change propels the material off the white dwarf's surface, a phenomenon known as 'thermonuclear runaway' \citep{1974ApJS...28..247S,1986ApJ...311..163S,1988ARA&A..26..377G,1989clno.conf...61B,Prialnik_1995,Starrfield_2009,starrfield2016thermonuclear,chomiuk2021new}, which gives rise to the nova outburst. 

During the nova's evolution, the bolometric luminosity remains relatively constant \citep{8}. The optical luminosity decreases as the peak intensity shifts from optical to UV to X-ray, tracing the optical light curve. However, the nova outburst is insufficient to destroy the white dwarf, implying the potential for recurrence \citep{1978ApJ...219..595F}. Based on this recurrence timescale, novae are classified as either classical (CNe) or recurrent (RNe). A nova is considered recurrent if the recurrence timescale ($\tau_{rec}$), the time between two successive outbursts, is less than 100 years; otherwise, it is classified as classical. The recurrence timescale of a nova is predominantly determined by the white dwarf mass ($M_{WD}$) and its mass accretion rate ($\dot{M}_{acc}$) \citep{1980A&A....85..295S,1982ApJ...253..798N,nomoto2007thermal}. Higher values of $M_{WD}$ and $\dot{M}_{acc}$ increase the likelihood of nova recurrence. This can be understood as follows: a higher white dwarf mass requires less material to trigger a thermonuclear runaway (TNR) explosion, and a higher accretion rate results in greater compressional heating and a shorter time to accumulate the critical ignition mass, leading to a shorter recurrence timescale. In addition to these theoretical insights, RNe also exhibit distinct observational features \citep{Duerback_1987,kato2003high,starrfield2012hydrodynamic,10.1093/mnras/stt1565,Schaefer2022} that differentiate them from CNe. However, none of these methods are very efficient in identifying RNe just from a single outburst.

Among the distinguishing features, the light curve of a nova provides valuable insights into the system. Optical light curves during an outburst reach a peak and gradually fade over days, weeks, or months, displaying specific shapes. These patterns classify novae into seven distinct categories: S (smooth decline), P (a plateau following a steep decline), D (a sudden dip), J (jitters), O (small oscillations due to a magnetic field), F (flat-topped), and C (a slow-rising, fast-falling cusp around the transition phase) \citep{6}. Notably, P-class novae have a strong association with recurrent novae (RNe), with 9 out of 11 identified RNe falling into this category. But there are also additional 22 P-class novae which are still known to be classical novae (CNe) \citep{Schaefer2022}. Statistical analysis also shows that 55.6\% – 88.9\% of RNe should belong to the P class, while only 16.9\% of CNe are to be classified as P class. These differences are statistically significant at the 8.8$\sigma$ level, considering binomial uncertainties, and at the 2100:1 level based on the Kolmogorov-Smirnov analysis \citep{pagnotta2014identifying}. Therefore, the presence of a P-class light curve is one of the important factors for distinguishing CNe from RNe. \citet{hachisu2009optical} assert that a "true plateau" is exclusive to RNe. However, it is suggested that a plateau can also occur when a broad bandpass includes emission line fluxes, causing the flux to remain nearly constant (continuum-plus-line flux) \citep{schaefer2010comprehensive}. This "false plateau" can occur in both RNe and CNe. A theoretical model proposed by \cite{kato1994optically}, based on their optically thick wind theory, explains the plateau signature in optical light curves, considering irradiation from the companion star and reprocessing of super soft X-ray emission from the white dwarf by the accretion disk. Their model successfully reproduces some of the observed P-class light curves \citep{hachisu1999theoretical,hachisu2001recurrent,hachisu2003revised}.

In conclusion, the presence of a plateau serves as a reasonable indicator of an RN. But there exist, to the best of our knowledge, no technique to distinguish RNe from CNe for P-class novae based solely on the observed light curve characteristics. Hence, there is a need for a technique capable of distinguishing CNe from RNe based solely on the characteristics of the P-class light curve. In the subsequent sections, we present a phenomenological approach to address this challenge.

\section{Sample selection}

For this work, we have collected well-sampled light curve data for all known Galactic P-class novae. These novae were selected based on the classification provided by \citet{Schaefer2022}. We note that the recurrent novae T CrB and V3890 Sgr exhibit S-class light curves, while V394 CrA, although classified as P-class, suffers from poor observational coverage. Consequently, these 3 RNe sources were excluded from our analysis, which focuses on the remaining 8 RNe with adequate light curve data. Similarly, out of 25 known P-class CNe, we are considering a subsample of 22 sources based on availability of light curves with good coverage. For further analysis, we have been taken the relevant physical parameters and distances from existing studies, list of which are summarized in the Table 1. The light curve data have been extracted from the AAVSO database, comprising magnitude values observed over specific periods, typically spanning months or years. In many instances, only visual (VIS) magnitude data were available, especially from historical records. To ensure consistency, we utilized only the VIS magnitude data for our analysis. However, V band magnitude can be transformed to VIS magnitude using the relation from \citet{Stanton1999}. Hence, we use both magnitudes depending on availability. For one specific light curve, of V2860 Ori's, visual data were not available. In that case, we used V-band and B-band data to calculate the corresponding VIS magnitudes, ensuring its inclusion in our analysis. Furthermore, the post-plateau phase for KT Eri was not entirely captured in the AAVSO database. So, we have instead used the V and B band data from \citet{9} for our work.

\begin{table}[h]
\centering
\caption{Novae samples for our analysis having well-sampled AAVSO data. All distances are from \cite{Schaefer2022}. Superscripts represent resources used for mass and accretion rate: (a) \cite{shara2018}, (b) \cite{10.1093/mnras/stt1565}, (c) \cite{hachisu2023multiwavelength}, (d) \cite{2010AJ....140.1347H}, (e) \cite{2008ApJ...687.1236H}, (f) \cite{2024ApJ...965...49H}, (g)\cite{2019ApJS..242...18H}. The $*$ on the year of outburts of KT Eri suggests these outbursts have been missed \citep{9}}.
\fontsize{7pt}{7pt}\selectfont
\begin{tabular}{c c c c c}
\hline
\multicolumn{5}{c}{\textbf{Classical Novae}} \\
\hline
Name & Year of First Outburst & Distance (pc) & $M_{WD}(M_\odot)$ & $log(\dot{M}_{acc}) (M_\odot /yr)$ \\ [0.8ex]
\hline
$\textnormal{DN Gem}^{(a)}$ & 1912 & 1381 & 1.14 & -9.24 \\
$\textnormal{CP Pup}^{(b)}$ & 1942 & 783 & 0.80 & -9.79 to -9.48 \\
$\textnormal{V1229 Aql}^{(a)}$ & 1970 & 3317 & 1.12 & -9.03 \\
$\textnormal{V368 Sct}$ & 1970 & 2452 & -- & -- \\
$\textnormal{HS Sge}^{(a)}$ & 1977 & 3193 & 1.13 & -9.89 \\
$\textnormal{V4021 Sgr}^{(a)}$ & 1977 & 8177 & 0.98 & -7.99 \\
$\textnormal{GQ Mus}^{(e)}$ & 1983 & 4043 & 0.7 & -- \\
$\textnormal{QU Vul}^{(a)}$ & 1985 & 1742 & 1.04 & -8.77 \\
$\textnormal{V444 Sct}$ & 1991 & 7381 & -- & -- \\
$\textnormal{V838 Her}^{(a)}$ & 1991 & 3128 & 1.35 & -10.02 \\
$\textnormal{V1974 Cyg}^{(a)}$ & 1992 & 1618 & 1.12 & -9.5 \\
$\textnormal{V351 Pup}^{(a)}$ & 1992 & 3869 & 1.19 & -9.76 \\
$\textnormal{BY Cir}^{(a)}$ & 1995 & 2396 & 1.04 & -8.6 \\
$\textnormal{V4633 Sgr}^{(a)}$ & 1998 & 2426 & 1.13 & -8.94 \\
$\textnormal{DD Cir}^{(a)}$ & 1999 & 3244 & 1.24 & -9.5 \\
$\textnormal{LZ Mus}^{(a)}$ & 1999 & 6749 & 1.27 & -8.16 \\
$\textnormal{V1065 Cen}^{(d)}$ & 2007 & 3322 & -- & -7.69 to -6.43 \\
$\textnormal{V339 Del}^{(f)}$ & 2007 & 1587 & 1.25 & -8.5 \\
$\textnormal{V1368 Cen}^{(g)}$ & 2012 & 6345 & 0.95 & -- \\
$\textnormal{V1535 Sco}$ & 2015 & 7790 & -- & -- \\
$\textnormal{V2860 Ori}$ & 2019 & 5969 & -- & -- \\
$\textnormal{YZ Ret}^{(c)}$ & 2020 & 2388 & 1.33 & -8.69 to -8.522 \\
\hline
\multicolumn{5}{c}{\textbf{Recurrent Novae}} \\
\hline
Name & Years of Outburst & Distance (pc) & $M_{WD}(M_\odot)$ & $log(\dot{M}_{acc}) (M_\odot /yr)$ \\ [0.8ex]
\hline
$\textnormal{U Sco}^{(a)}$ & 1863, 1906, 1936, 1979, 1987, 1999, 2010, 2022 & 6258 & 1.36 & -7.29 \\
$\textnormal{KT Eri}^{(a)}$ & $1889^{*}$, $1919^{*}$, $1949^{*}$, $1979^{*}$, 2009 & 4211 & 1.25 & -6.46 \\
$\textnormal{V745 Sco}^{(a)}$ & 1897, 1937, 1989, 2014 & 8016 & 1.40 & -7.96 \\
$\textnormal{RS Oph}^{(a)}$ & 1898, 1907, 1933, 1958, 1967, 1985, 2006, 2021 & 2710 & 1.31 & -7.14 \\
$\textnormal{V2487 Oph}^{(a)}$ & 1900, 1998 & 7539 & 1.31 & -7.14 \\
$\textnormal{CI Aql}^{(a)}$ & 1917, 1941, 2000 & 2761 & 1.21 & -6.95 \\
$\textnormal{IM Nor}^{(a)}$ & 1920, 2002 & 4312 & 1.21 & -7.32 \\
$\textnormal{T Pyx}^{(a)}$ & 1968, 2011 & 3599 & 1.23 & -6.95 \\

\hline
\end{tabular}
\end{table}

\section{Data analysis}

First we use all available data to produce a light curve by computing mean magnitude with one day binning for each Julian Day. Figure 1 shows the example light curve for the 1992 outburst of V1974 Cyg. The time is shown in days from the outburst peak (occuring at t = $T_0$), and the aparent VIS magnitude is after one day binning.

\subsection{Parameterizing the light curve}

Figure 1 illustrates the P-class light curve, showcasing a steep decay phase lasting for almost 2 months, followed by the plateau phase with a relatively gentler slope. The transition point from the pre-plateau to the plateau phase is termed the 'first breakpoint' ($T_1$), while the onset of the decay with a steeper slope than the plateau phase marks the beginning of the post-plateau phase, denoted as the 'second breakpoint' ($T_2$). Subsequently, the light curve progresses towards the quiescent phase. We aim to identify suitable parameters that distinctly characterize each nova light curve. We use the light curve of V1974 Cyg as a reference and aim to find parameters for all other light curves (test LCs) that transform them into our reference LC. For recurrent novae (RNe), we consider the light curves for each of their outbursts separately.

We identify six parameters essential to distinctly describe a nova light curve: initial magnitude offset ($\Delta m$), time scaling factors for the pre-plateau ($\alpha_1$) and plateau phases ($\alpha_2$), and the rates of decline for the pre-plateau ($m_1$), plateau ($m_2$), and post-plateau ($m_3$) phases. The initial magnitude offset $\Delta m$ is the difference between the intercepts of the pre-plateau phase for the reference and test light curves. The time scaling factors are computed as:
\[
\alpha_1 = \frac{T_{1,t} - T_{0,t}}{T_{1,\text{ref}} - T_{0,\text{ref}}}
\]
and
\[
\alpha_2 = \frac{T_{2,t} - T_{1,t}}{T_{2,\text{ref}} - T_{1,\text{ref}}}
\]
where $T_{0,t}$, $T_{1,t}$, and $T_{2,t}$ are the breakpoints for the test light curve, and $T_{0,\text{ref}}$, $T_{1,\text{ref}}$, and $T_{2,\text{ref}}$ are the corresponding breakpoints for the reference light curve. The parameters $m_1$, $m_2$, and $m_3$ represent the slopes for each segment. The values of these parameters for our nova light curves are provided in Table 2. To segment the P-class light curve into three distinct evolutionary phases, we utilize the \texttt{PiecewiseLinFit} algorithm from the \texttt{pwlf} Python library. This data-driven approach allows for robust identification of breakpoints and fitting of a piecewise linear model with three segments to the observed light curve. Following the determination of these breakpoints, we employ \texttt{polyfit} function from \texttt{numpy} to estimate the uncertainties in the corresponding slopes and intercepts. The automated nature of this methodology ensures consistency across the dataset. In cases where the inferred breakpoints significantly diverged from those reported in the literature—likely due to data sparsity or irregularities in the light curve—we adopted literature-based values \citep{6} to maintain physical credibility. Representative fits illustrating the effectiveness of this method are provided in the appendix.

\begin{figure}[h]
  \centering
  \includegraphics[width=0.75\textwidth, , height= 6.5 cm]{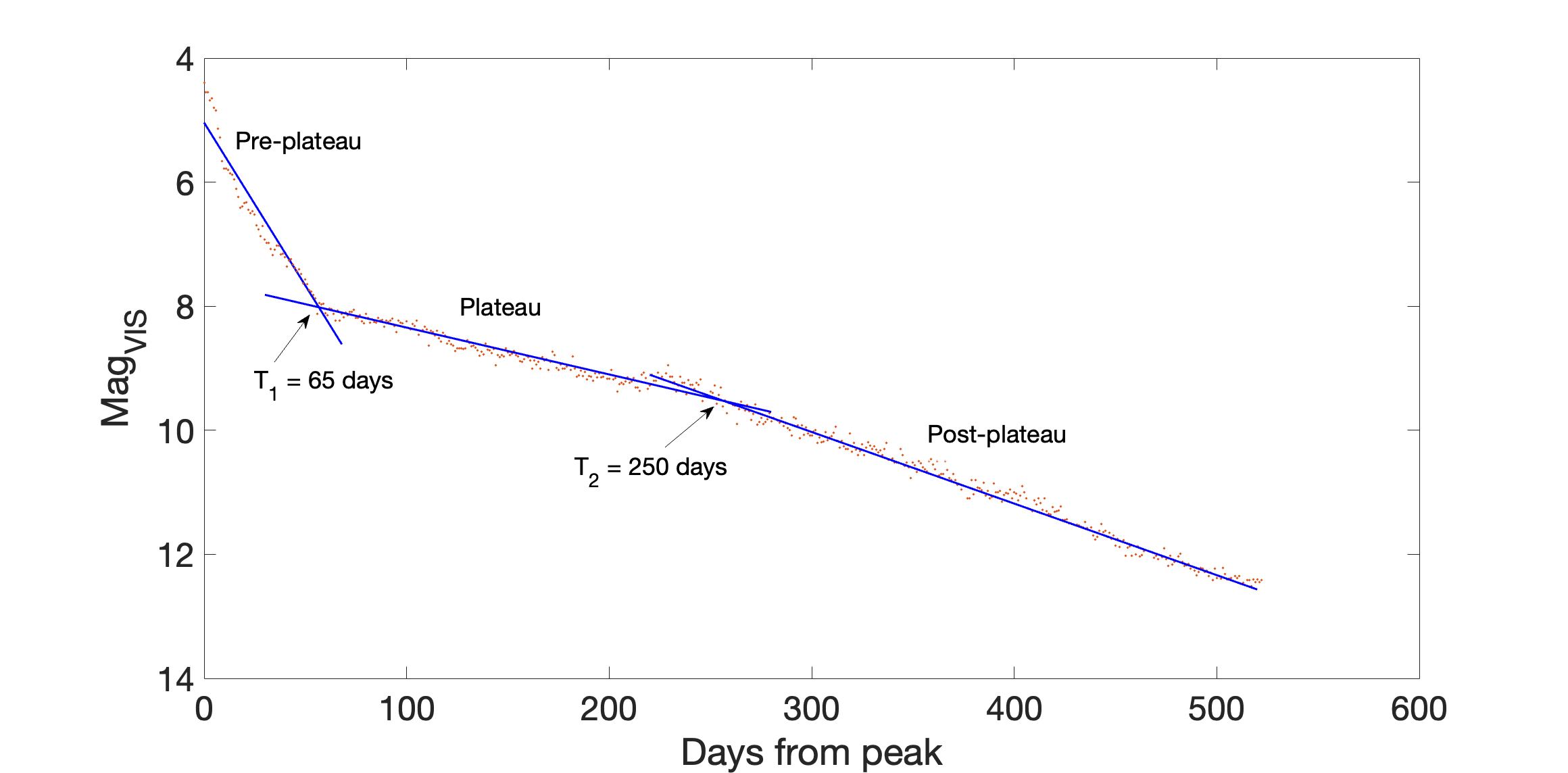}
  \caption{The optical light curve of V1974 Cyg (taken as reference). We have shown three different phases: pre-plateau, plateau and post-plateau phases the corresponding linear fits.}
\end{figure}

\subsection{Clustering and correlation of parameters}

Our principal objective is to investigate the potential role of these six key parameters in distinguishing recurrent novae (RNe) from classical novae (CNe). We check scatter plot for all different combinations of parameters to see if clustering and separation of parameters for RNe and CNe can be identified. We also tried to identify if a combination of these parameters can be used more effectively to separate RNe and CNe. 

Subsequently, we also aim to provide a theoretical foundation for our empirical observations. From the Nomoto plot (Figure 4 in the \citet{nomoto2007thermal}), it is evident that white dwarf mass ($M_{WD}$) and mass accretion rate ($\dot{M}_{acc}$) are related to a nova's recurrent nature and the related timescale. Therefore, we investigate potential correlations between our set of six observational parameters and these crucial physical attributes ($M_{WD}$ and $\dot{M}_{acc}$). To ensure uniformity in finding the correlations, we exclusively use the novae mass and accretion rates from \citet{shara2018}.

We compute Pearson's correlation coefficient and p-values for each combination of the observational parameters and the physical parameters of the white dwarf. The Pearson coefficient ($\alpha$) measures the strength and direction of the linear relationship between two variables, ranging from -1 to 1. The corresponding p-value reveals the significance of the correlation. The $\alpha$ and p-values of the light curve parameters with that of $M_{WD}$ and $\dot{M}_{acc}$ are listed in Table 3.

\begin{table}[h]
\centering
\caption{Best values of parameters for all P-class novae}
\fontsize{9pt}{9pt}\selectfont
\begin{tabular}{c c c c c c c} 
\toprule
\multicolumn{7}{c}{\textbf{Classical Novae}} \\
\toprule
Name & $\Delta m$ & $\alpha_1$ & $\alpha_2$ & $m_1$ & $m_2$ & $m_3$ \\ [0.8 ex] 
\midrule
V1974 Cyg & 0 & 1 $\pm$ 0.0399 & 1 $\pm$ 0.0352 & 0.0526 $\pm$ 0.0017  & 0.0078 $\pm$ 0.0003 & 0.0115 $\pm$ 0.0001\\ 	
QU Vul & 1.076 $\pm$ 0.0035 & 1.2 $\pm$ 0.0512 & 3.338 $\pm$ 0.0847 & 0.0572 $\pm$ 0.0018 & 0.0021 $\pm$ 0.00006 & 0.0039 $\pm$ 0.00005\\
DN Gem & -0.032 $\pm$ 0.0034 & 1.03 $\pm$  0.0706 & 1.30 $\pm$ 0.1186	 & 0.0378 $\pm$ 0.0017 & 0.0024 $\pm$ 0.0004 & 0.0072 $\pm$ 0.0002\\
CP Pup & -3.174 $\pm$ 0.0148 & 0.415 $\pm$ 0.0535 & 0.525 $\pm$ 0.0919 & 0.159 $\pm$ 0.013	& 0.0032 $\pm$ 0.0008	& 0.0104 $\pm$ 0.0007\\
BY Cir & 3.785 $\pm$ 0.0022  & 2.615 $\pm$ 0.3554 & 0.781 $\pm$ 0.2889	 & 0.0138 $\pm$ 0.0005 & 0.0019 $\pm$ 0.0013	& 0.014 $\pm$ 0.001\\ 
HS Sge & 3.533 $\pm$ 0.0081 & 0.707 $\pm$ 0.0214 & 3.028 $\pm$ 	0.0325	& 0.0667 $\pm$ 0.0012 & 0.0034 $\pm$ 0.0002	 & 0.0129 $\pm$ 0.0004\\
V838 Her & 4.586 $\pm$ 0.0066  & 0.846 $\pm$ 0.07902 & $>$ 1.143 & 0.1055 $\pm$ 	0.0074	& 0.0013 $\pm$ 	0.001 & none\\ 	
V1535 Sco & 4.656 $\pm$ 0.0068  & 0.446 $\pm$ 0.0182 & 0.388 $\pm$	0.0635 & 0.1322 $\pm$ 	0.0035	 & 0.009 $\pm$ 0.001	& 0.0136 $\pm$ 0.0006\\ 	
V2860 Ori & 7.044 $\pm$  0.0045 & 1.20 $\pm$ 0.1355 & 0.468 $\pm$ 0.1834 & 0.0585 $\pm$ 0.0025	& 0.0078 $\pm$ 0.0031 & 0.0237 $\pm$ 0.0021\\
V4021 Sgr & 4.12 $\pm$ 0.0285	& 0.892 $\pm$  0.0518 & 1.6 $\pm$ 0.3480 & 0.0383 $\pm$ 0.0015 & 0.0037 $\pm$ 0.0006	& 0.014 $\pm$ 0.0018 \\
V4633 Sgr & 3.668 $\pm$ 0.0343	& 1.169 $\pm$  0.0714 & 1.238 $\pm$ 0.1725 & 0.0384 $\pm$ 0.0017	& 0.0017 $\pm$ 0.0004	& 0.005 $\pm$ 0.0002\\ 
V1065 Cen & 3.777 $\pm$ 0.0094	& 0.477 $\pm$ 0.0492 & 0.206 $\pm$ 0.1125	& 0.0968 $\pm$ 0.0045	 & 0.0031 $\pm$ 0.0041	& 0.0182 $\pm$ 0.0008\\
GQ Mus & 3.214 $\pm$ 0.0051 & 0.953 $\pm$ 0.1775 & 2.625 $\pm$ 0.2887 & 0.0340 $\pm$ 0.0057 & 0.0009 $\pm$ 0.0005 & 0.0063 $\pm$ 0.0002\\
YZ Ret & -0.46 $\pm$ 0.0027 & 0.338 $\pm$  0.0235 & 0.125 $\pm$ 0.0350 & 0.1374 $\pm$ 0.0076 & 0.0085 $\pm$ 0.0029 & 0.0244 $\pm$ 0.0008\\
V368 Sct & 3.123 $\pm$ 0.0062 & 0.646 $\pm$  0.0359 & $>$ 12  & 0.0723 $\pm$ 	0.0033 & 0.0027 $\pm$ 0.0002 & none\\
LZ Mus & 5.564 $\pm$ 0.0218 & 0.492 $\pm$ 0.0379 & 0.606 $\pm$ 0.0521 & 0.0989 $\pm$ 0.005	& 0.0112 $\pm$ 0.001 & 0.0194 $\pm$ 0.001\\
V1229 Aql & 2.212 $\pm$ 0.0029 & 1.154 $\pm$ 0.0287 & $>$ 26.125 & 0.0928 $\pm$ 0.0027	& 0.0004 $\pm$ 0.00006 & none\\
DD Cir & 4.145 $\pm$ 0.0033	 & 0.307 $\pm$  0.0669 & $>$ 0.1375 & 0.3122 $\pm$ 0.02 & 0.0464 $\pm$ 0.004 & none\\
V351 Pup & 3.856 $\pm$ 0.0035 & 0.969 $\pm$  0.0572 & 2.006 $\pm$ 0.2695 & 0.0287 $\pm$ 0.0011 & 0.005 $\pm$ 0.0004 & 0.0102 $\pm$ 0.0011\\
V444 Sct & 6.16 $\pm$ 0.045 & 0.185	$\pm$  0.0307 & $>$ 0.162 & 0.212 $\pm$ 0.026	& 0.025 $\pm$ 0.019	& none\\
V339 Del & 0.12	$\pm$ 0.256 & 1.046 $\pm$ 0.0769 & 0.468 $\pm$ 0.29	& 0.0853 $\pm$ 0.006 & 0.0029 $\pm$ 0.0006	 & 0.0056 $\pm$ 0.0003\\
V1368 Cen & 4.84 $\pm$ 2.76 & 0.692 $\pm$ 0.0308 & $>$ 0.3188 & 0.0825 $\pm$ 	0.0026 & 0.0146 $\pm$ 0.001 & none\\
\bottomrule
\multicolumn{7}{c}{\textbf{Recurrent Novae}} \\
\toprule
Name & $\Delta m$ & $\alpha_1$ & $\alpha_2$ & $m_1$ & $m_2$ & $m_3$ \\ [0.8 ex] 
\midrule
$\textnormal{CI Aql (2000)}$ & -0.2 $\pm$ 0.0092 & 1.123 $\pm$ 0.0605 & $>$ 0.868 & 0.0515 $\pm$ 	0.002 & 0.0055 $\pm$ 0.0008 & none\\
$\textnormal{IM Nor (2002)}$ & 3.746 $\pm$ 0.0058 & 1.723 $\pm$ 0.0959  & $>$ 0.80 & 0.0342 $\pm$ 0.0007	& 0.0086 $\pm$ 	0.0012 & none\\
$\textnormal{RS Oph (1958)}$ & 1.382 $\pm$ 0.0056 & 0.523 $\pm$ 0.0409 & 0.25 $\pm$ 0.0456 & 0.0921 $\pm$ 0.004 & 0.0254 $\pm$ 0.002 & 0.034 $\pm$ 0.0025\\
$\textnormal{RS Oph (1967)}$ & 0.986 $\pm$ 0.0065 & 0.492 $\pm$ 0.0324 & 0.306 $\pm$ 0.0574 & 0.1034 $\pm$ 0.0047	& $<$ 0.018  & 0.0264 $\pm$ 0.0021\\
$\textnormal{RS Oph (1985)}$ & 0.991 $\pm$  0.0075	& 0.492 $\pm$  0.0385 & 0.275 $\pm$ 0.0362 & 0.0811 $\pm$ 0.0048	& 0.0316 $\pm$ 0.0014 & 0.0453 $\pm$ 0.0022\\
$\textnormal{RS Oph (2006)}$ & 1.367 $\pm$ 0.0053 & 0.646 $\pm$ 0.0494 & 0.2125 $\pm$ 0.0393	& 0.0836 $\pm$ 0.0039 & 0.0095 $\pm$ 0.0015 & 0.0469 $\pm$ 	0.0015\\
$\textnormal{RS Oph (2021)}$ & 1.233 $\pm$ 0.0065 & 0.415 $\pm$ 0.0303 & 0.306 $\pm$ 0.0342	& 0.1198 $\pm$ 0.0044 & 0.0246 $\pm$ 0.0017 & 0.0535 $\pm$ 	0.0019\\
$\textnormal{KT Eri (2009)}$ & 2.75	$\pm$ 0.0043	& 1.308 $\pm$ 0.0923 & 0.975 $\pm$ 0.25	& 0.039 $\pm$ 0.001	& 0.0057 $\pm$ 0.0029	& 0.038 $\pm$ 0.0012\\
$\textnormal{U Sco (2010)}$ & 3.54 $\pm$ 0.0063 & 0.154 $\pm$ 0.0099 & 0.119 $\pm$ 0.0157 & 0.503 $\pm$ 0.0267 & 0.0473  $\pm$ 0.0047	& 0.1889 $\pm$ 0.0108\\
$\textnormal{U Sco (2022)}$ & 4.097 $\pm$ 0.0074 & 0.155 $\pm$ 0.0161	& 0.094 $\pm$ 0.0165 & 0.459 $\pm$ 	0.0325	& 0.056 $\pm$ 0.0054	& 0.133 $\pm$ 0.0112\\
$\textnormal{T Pyx (1968)}$ & 2.135 $\pm$ 0.0027  & 1.615 $\pm$ 0.0683 & 0.50 $\pm$ 0.1291 & 0.0342 $\pm$ 0.0009 & 0.01 $\pm$ 0.0005	& 0.02 $\pm$ 0.0021\\
$\textnormal{T Pyx (2011)}$ & 1.782 $\pm$ 0.0035 & 1.8 $\pm$ 0.0709 & 0.50 $\pm$ 0.1405	& 0.0367 $\pm$ 0.0009 & 0.0105 $\pm$ 0.0005 & 0.02 $\pm$ 0.0019\\
$\textnormal{V2487 Oph (1998)}$ & 5.524 $\pm$ 0.0268 & 0.123 $\pm$ 0.0133 & 0.094 $\pm$ 0.0237 & 0.3468 $\pm$ 0.025 & 0.0358 $\pm$ 0.008	& 0.0674 $\pm$ 0.0024\\
$\textnormal{V745 Sco (2014)}$ & 4.662 $\pm$ 0.0222 & 0.246 $\pm$  0.0195	& $>$ 0.131 & 0.2631 $\pm$ 0.0204 & 0.0464 $\pm$ 	0.004 & none\\
\bottomrule
\end{tabular}
\end{table}

\section{Results}
\subsection{The connection}

The analysis of the white dwarf mass ($M_{\text{WD}}$) reveals strong correlations with several observational parameters, confirming previous findings that a higher $M_{\text{WD}}$ corresponds to a more rapid decline \citep{Yaron_2005}.  Our results indicate significant correlations between $M_{\text{WD}}$ and the parameters $m_1$, $m_2$, $m_3$, as well as the time-scaling factors $\alpha_1$ and $\alpha_2$. Specifically, $M_{\text{WD}}$ shows Pearson correlation coefficients of 0.68 ($p = 0.008$) for $m_1$, 0.76 ($p = 0.002$) for $m_2$, and 0.66 ($p = 0.01$) for $m_3$, indicating strong and significant positive correlations. Figure 2 (a) gives an idea of the existing correlation. Additionally, $M_{\text{WD}}$ correlates negatively with $\alpha_1$ and $\alpha_2$, with coefficients of -0.57 ($p = 0.03$) and -0.66 ($p = 0.01$), respectively. Here to note that \citet{shara2018} have not reported uncertainties for individual mass and accretion rate but have only mentioned possible typical uncertainty of 0.1$M_\odot$ for WD mass.

On the other hand, the mass accretion rate ($\log(\dot{M}_{\text{acc}})$) shows fewer significant correlations. It is negatively correlated with $\alpha_2$ (Pearson coefficient = -0.70, $p = 0.003$) and weakly positively correlated with $m_2$ (0.47, $p = 0.09$) and $m_3$ (0.46, $p = 0.10$). Although these correlations between $\log(\dot{M}_{\text{acc}})$ and $m_2$, $m_3$ are not very strong, they remain statistically significant. The negative correlation of $\alpha_2$ and $m_3$ with $\log(\dot{M}_{\text{acc}})$ could indicate complex interactions during the plateau and post-plateau phases, possibly involving accretion disk irradiation. Nevertheless, the underlying mechanism behind these correlations remains uncertain and requires further investigation.

\subsection{Separation of CNe and RNe}

From plots (b), (d) and (f) of Figure 2, it is evident that, the rate of decline for RNe is consistently higher across all three plots, as anticipated. However, the selected combinations of parameters are insufficient to completely separate classical novae (CNe) from recurrent novae (RNe). This is due to the strong correlation of these parameters with $M_{\text{WD}}$, while the weaker correlation with $\log(\dot{M}_{\text{acc}})$. A particularly noteworthy result is depicted in the Figure 2 (f), which shows the relationship between $m_3$ and $\alpha_2$ — the parameter with the strongest correlation with $\log(\dot{M}_{\text{acc}})$ among those analyzed. In this plot, a more distinct separation between CNe and RNe emerges. Specifically, classical novae exhibit a maximum cutoff value for $m_3$ and a minimum for $\alpha_2$, while RNe display the opposite trend. This pattern reflects the established correlations between the observational parameters and the underlying physical properties of the nova systems. The cutoff values of $m_3$ and $\alpha_2$ for CNe correspond to lower mass accretion rates and smaller white dwarf masses, as opposed to the higher values observed in RNe.

The complexity of the relationships among these parameters shows the inadequacy of a two-parameter analysis in achieving a clear distinction between CNe and RNe. To address this limitation, we incorporate all the available parameters that exhibit significant correlations with the physical characteristics of the systems, namely $\vec{x}$ = \{$\alpha_1$, $\alpha_2$, $m_1$, $m_2$, $m_3$\}. We model the separation using the following linear equation:

$$
\mathcal{F}(\vec{x}) = a_1 \alpha_1 + a_2 \alpha_2 + a_3 m_1 + a_4 m_2 + a_5 m_3 + a_6
$$

where the $a_i$'s represent the coefficients to be estimated. To determine these coefficients, we use the \texttt{LogisticRegression} model from the \texttt{sklearn} package in Python, with the \texttt{liblinear} solver. The resulting coefficients are $a_1 = 1.28$, $a_2 = -1.04$, $a_3 = -14.90$, $a_4 = 272.02$, $a_5 = 113.15$, and $a_6 = -5.05$.

Due to the relatively small size of our dataset, we employ a simple classifier model. Figure 3 presents the results of this hyperplane, where the \textit{rate of decline component} is given by:

$$
-14.90 m_1 + 272.02 m_2 + 113.15 m_3 
$$

and the \textit{time-scaling component} is represented as:

$$
- 1.28 \alpha_1 + 1.04 \alpha_2 + 5.05
$$

We also examined the potential for identifying evolutionary tracks in this parameter space, similar to recurrence timescale contours in the $M_{\rm WD}$–$\log(\dot{M}{\rm acc})$ plane. In principle, such tracks could emerge if our parameters could be reliably mapped to physical quantities like the white dwarf mass and accretion rate. While this remains challenging—particularly due to the limited sample size—we attempted a preliminary visual exploration. In the same Figure 3, for Recurrent Novae (RNe),we use marker sizes scaled by their recurrence timescales. A tentative trend appears: systems with longer recurrence timescales tend to lie toward lower values of both the rate-of-decline and time-scaling components. Although the pattern is intriguing, establishing firm evolutionary contours will require a larger and more complete sample of P-class novae. But this approach may serve as a promising starting point. We also note that V2860 Ori is lying at the very boundary of the CNe and RNe as shown in Figure 3. This is indicative of the possibility of its recurrence in near future and possible missing nova outburst in the past.

\begin{table}
\caption{Pearson coefficient and p-values for different parameter pairs. The first term in the bracket is Pearson's coefficient, and the second term is the corresponding p-value. It is important to note that $\Delta M$ represents the magnitude offset on an absolute scale derived from $\Delta m$ (We adopt visual extinction values, $A_V$, from the dust maps of \citet{1998ApJ...500..525S})}.
\centering
\fontsize{8pt}{8pt}\selectfont
\begin{tabular}{c c c c c} 
\hline
\noalign{\smallskip}
 Parameters & $M_{WD}$ & $\log(\dot M_{acc})$ \\ [0.5ex] 
\noalign{\smallskip}
\hline
\noalign{\smallskip}
$\Delta M$ & (0.24, 0.56) & (-0.15, 0.43)\\
\noalign{\smallskip}
\hline
\noalign{\smallskip}
$\alpha _1$ &  (-0.57, 0.03) & (-0.12, 0.66)\\
\noalign{\smallskip}
\hline
\noalign{\smallskip}
$\alpha _2$ & (-0.66, 0.01) & (-0.70, 0.003)\\
\noalign{\smallskip}
\hline
\noalign{\smallskip}
$m_1$ & (0.68, 0.008) & (0.37, 0.18)\\
\noalign{\smallskip}
\hline
\noalign{\smallskip}
$m_2$ & (0.76, 0.002) & (0.47, 0.09)\\
\noalign{\smallskip}
\hline
\noalign{\smallskip}
 $m_3$ & (0.66, 0.01) & (0.46, 0.10)\\
 \noalign{\smallskip}
\hline
\noalign{\smallskip}
\end{tabular}

\end{table}

\begin{figure}
  \centering
  \subfloat[]{\includegraphics[width=0.50\textwidth]{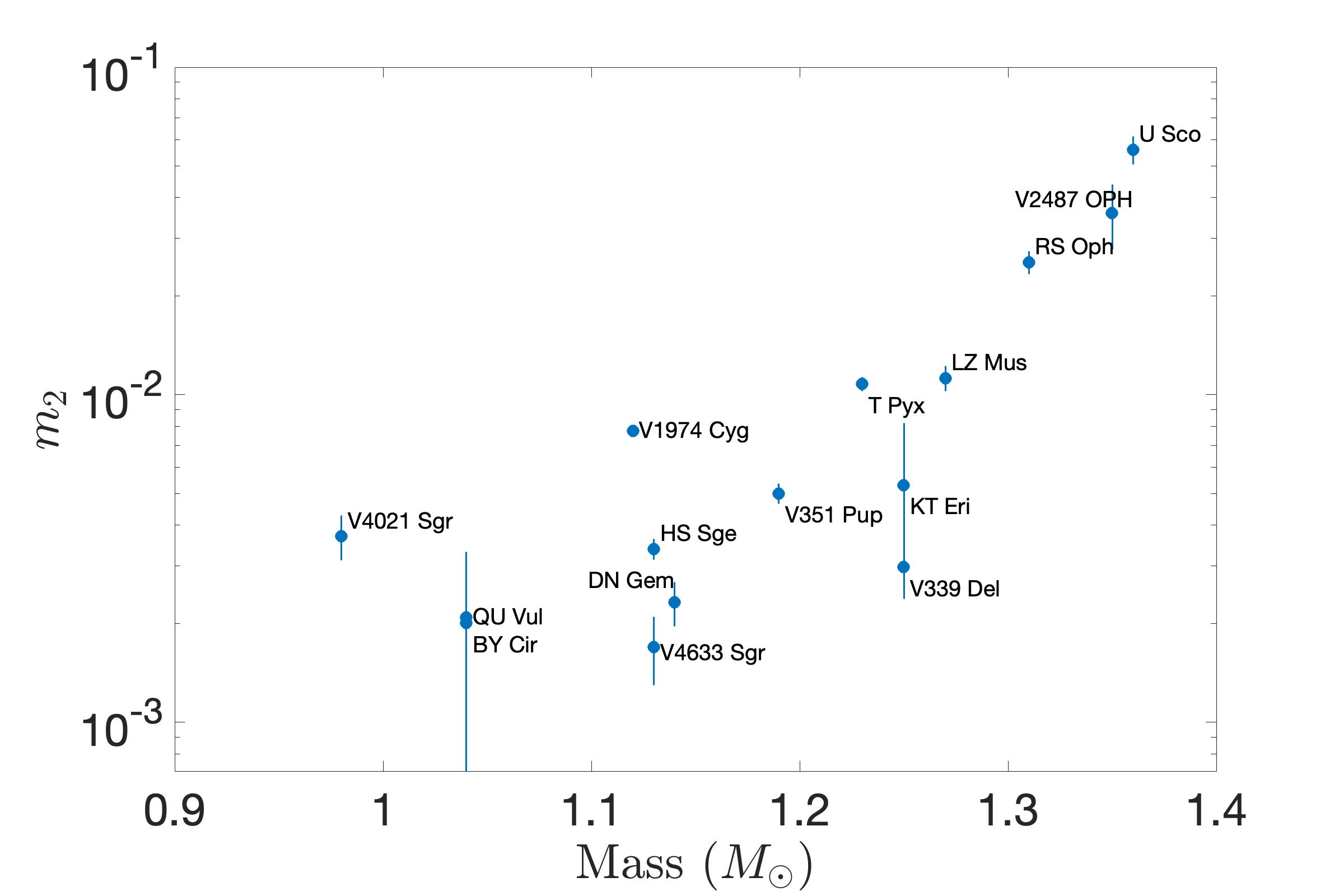}}\hfill
  \subfloat[]{\includegraphics[width=0.50\textwidth]{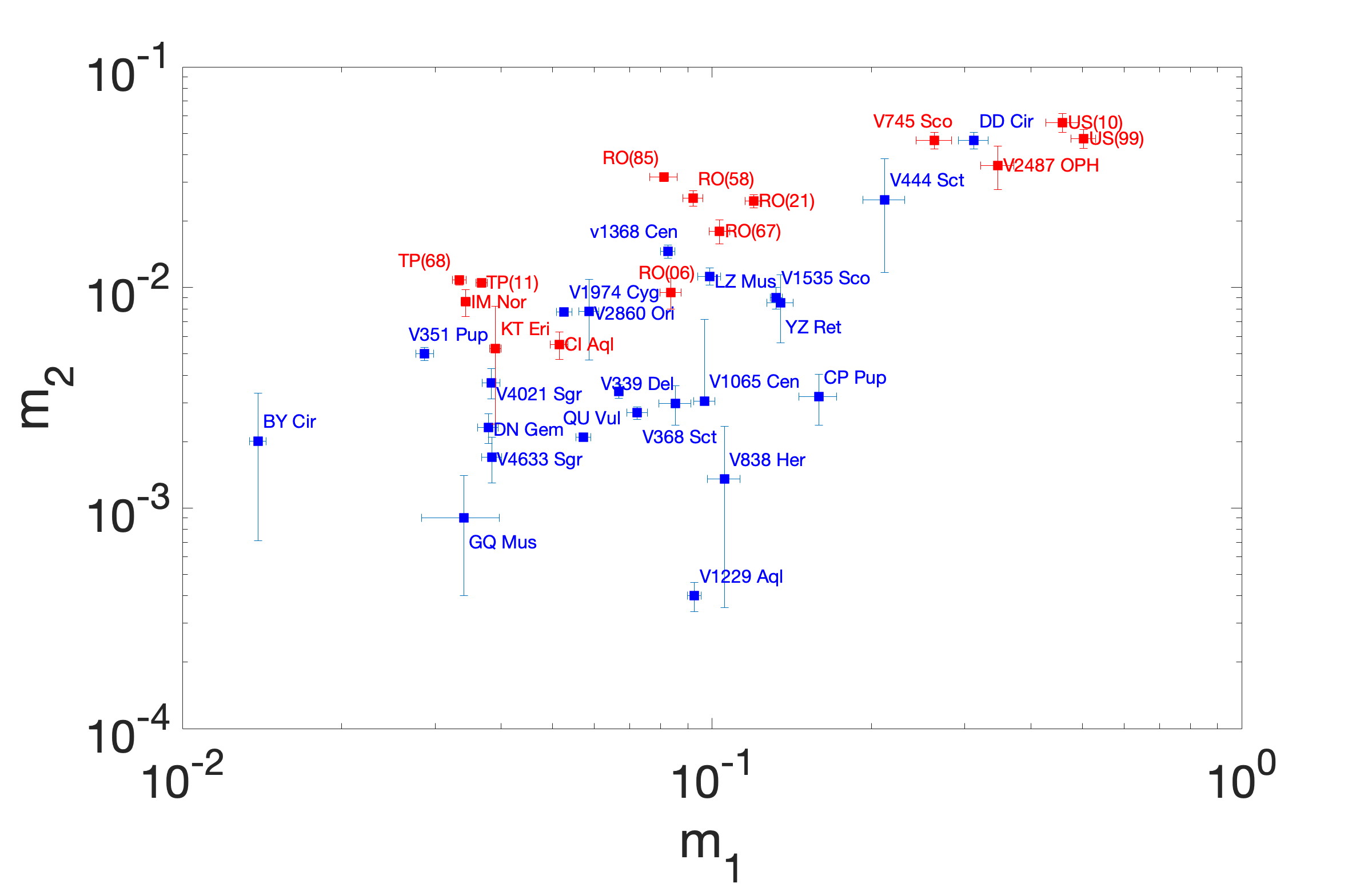}} \\
  \subfloat[]{\includegraphics[width=0.50\textwidth]{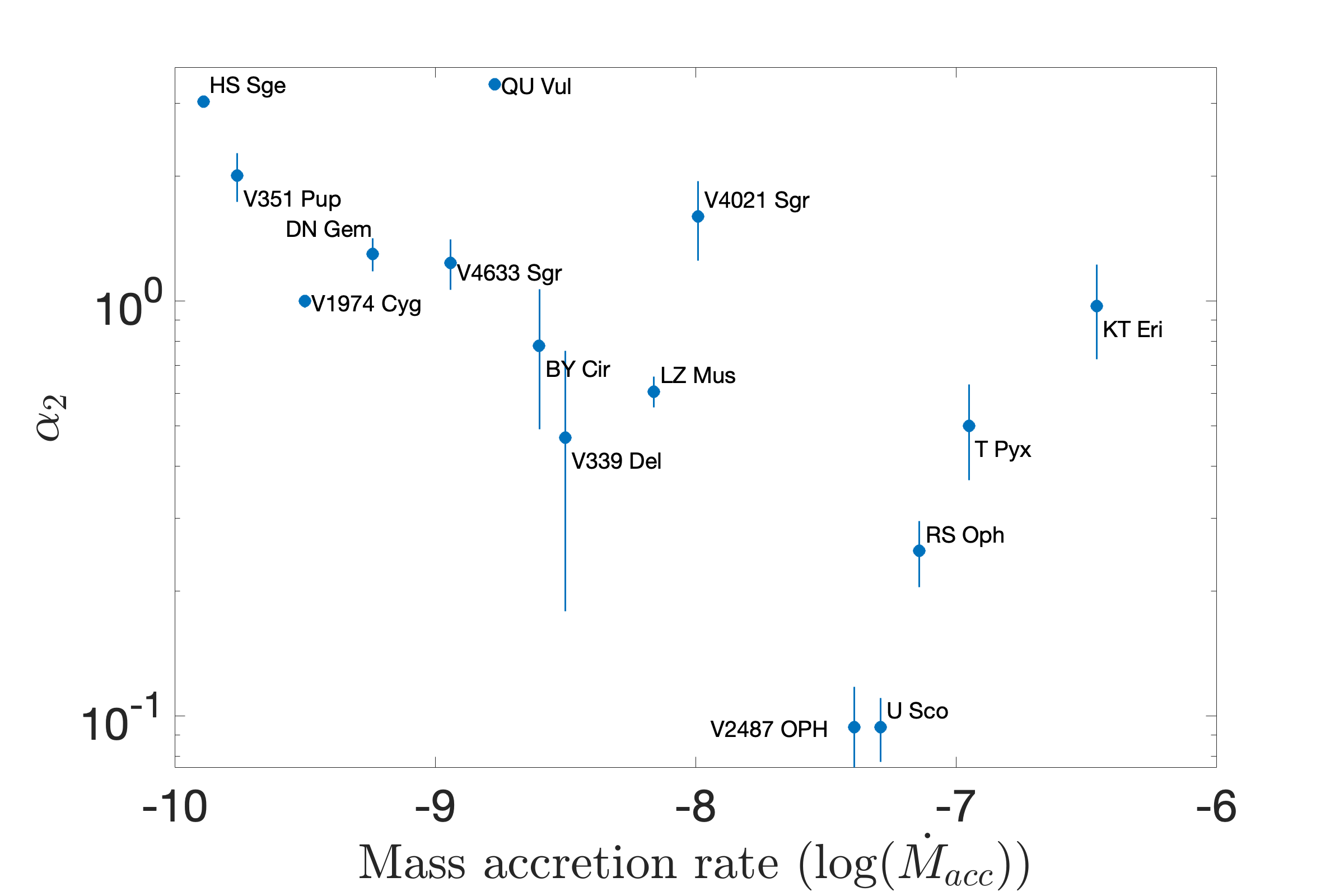}}\hfill
  \subfloat[]{\includegraphics[width=0.50\textwidth]{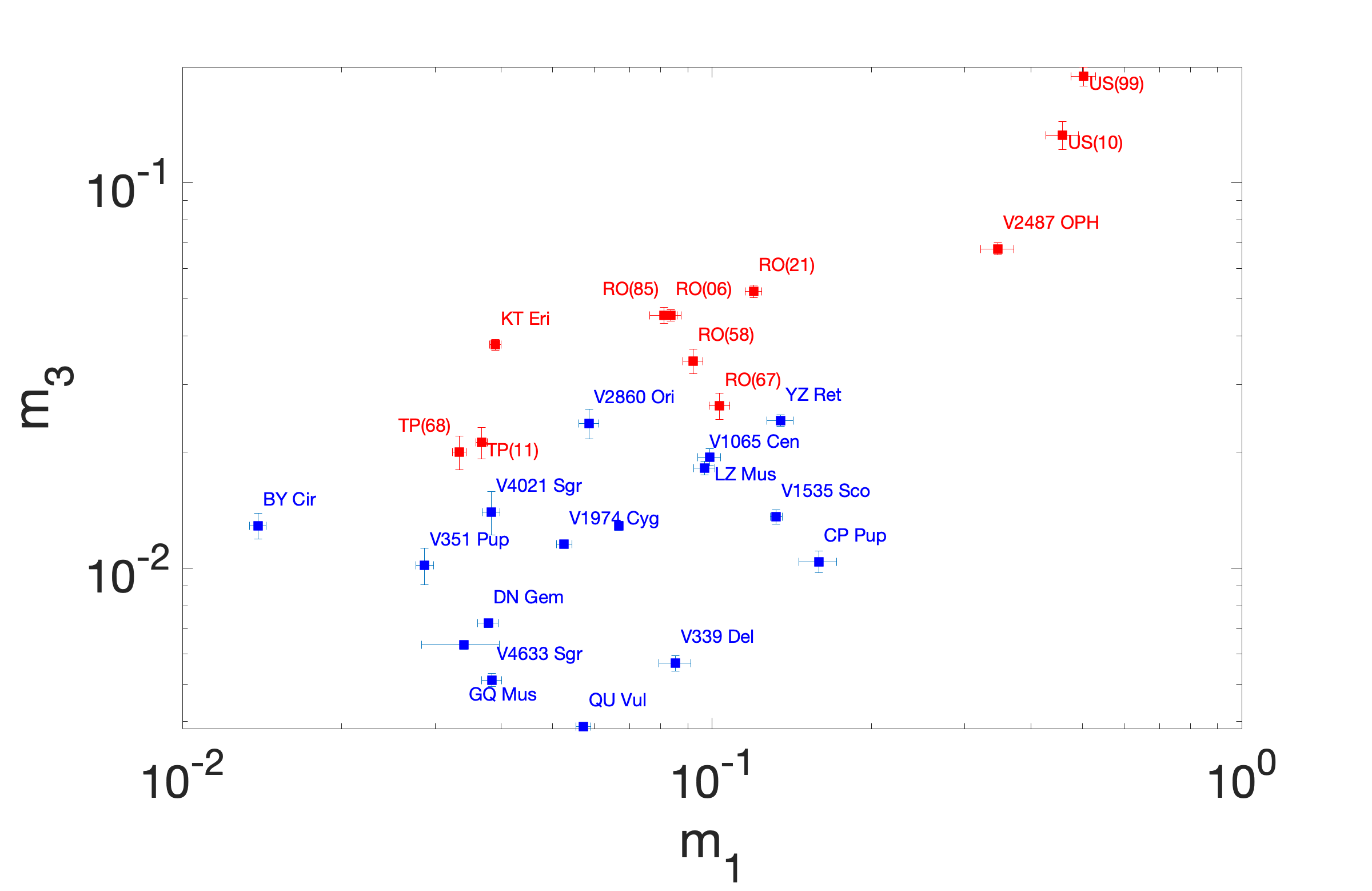}} \\
  \subfloat[]{\includegraphics[width=0.50\textwidth]{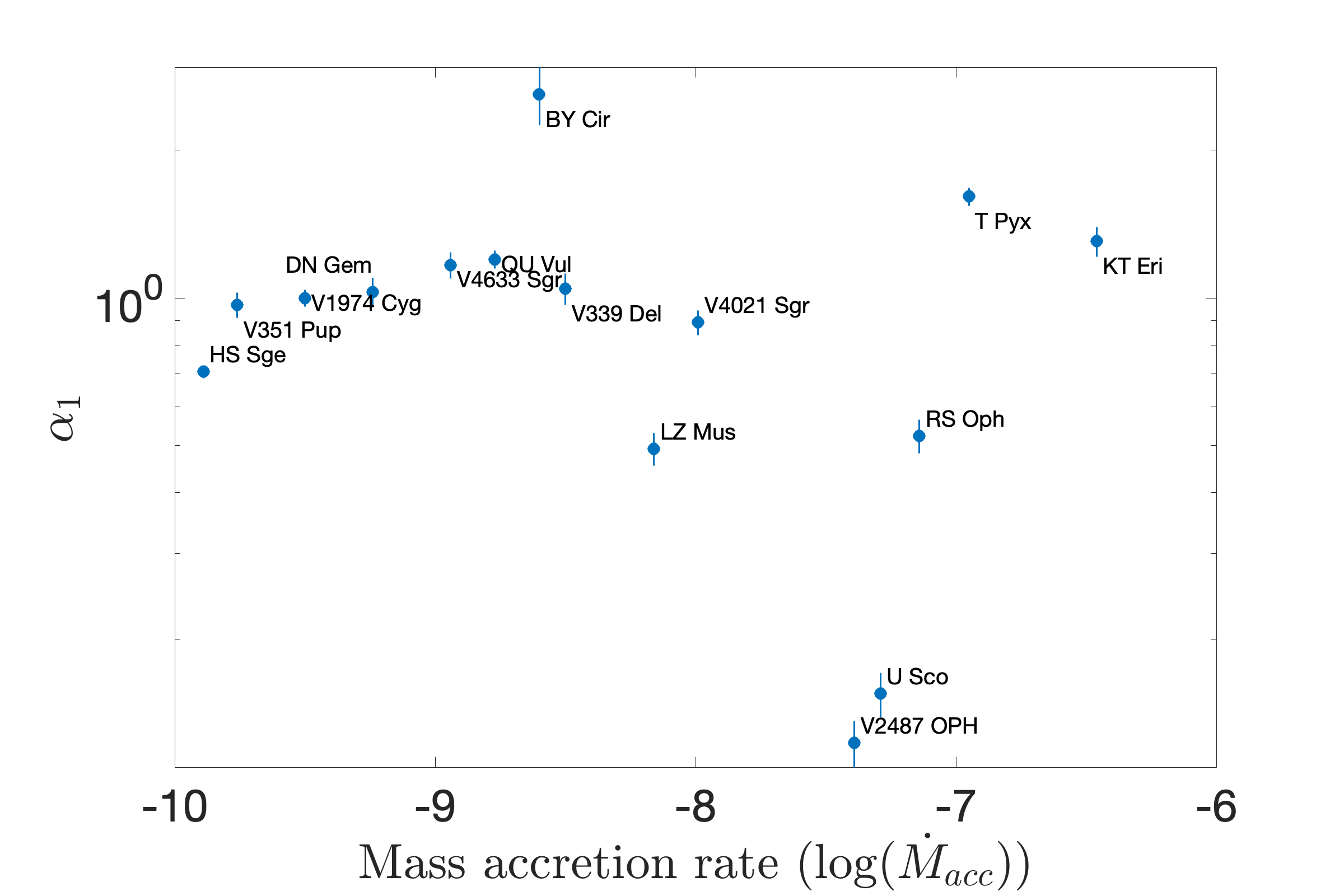}}\hfill
  \subfloat[]{\includegraphics[width=0.50\textwidth]{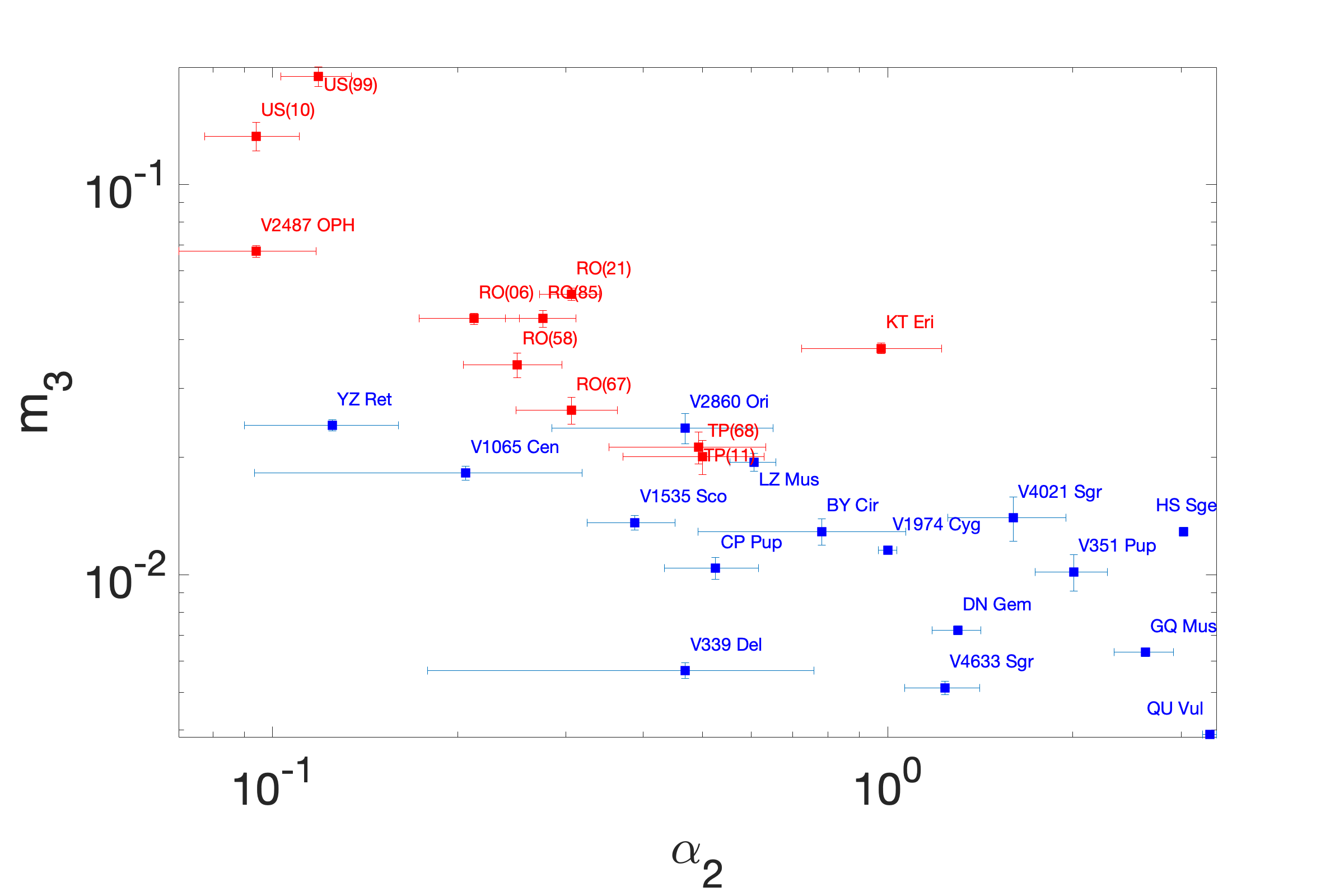}}
  \caption{\textit{Left column}: Relation between observed light curve parameters with physical parameters - white dwarf mass and the accretion rate - for the sample. Plots (a) and (c) show example of significant correlation between these parameters, while plot (e) is an example of no significant correlation. \textit{Right column}: Example of correlations among observed light curve parameters. RS Oph, T Pyx and U Sco are indicated as RO, TP, US, respectively.}
  \label{fig:relations}
\end{figure}

\begin{figure}
  \centering
  \includegraphics[width=1\textwidth, height=12cm]{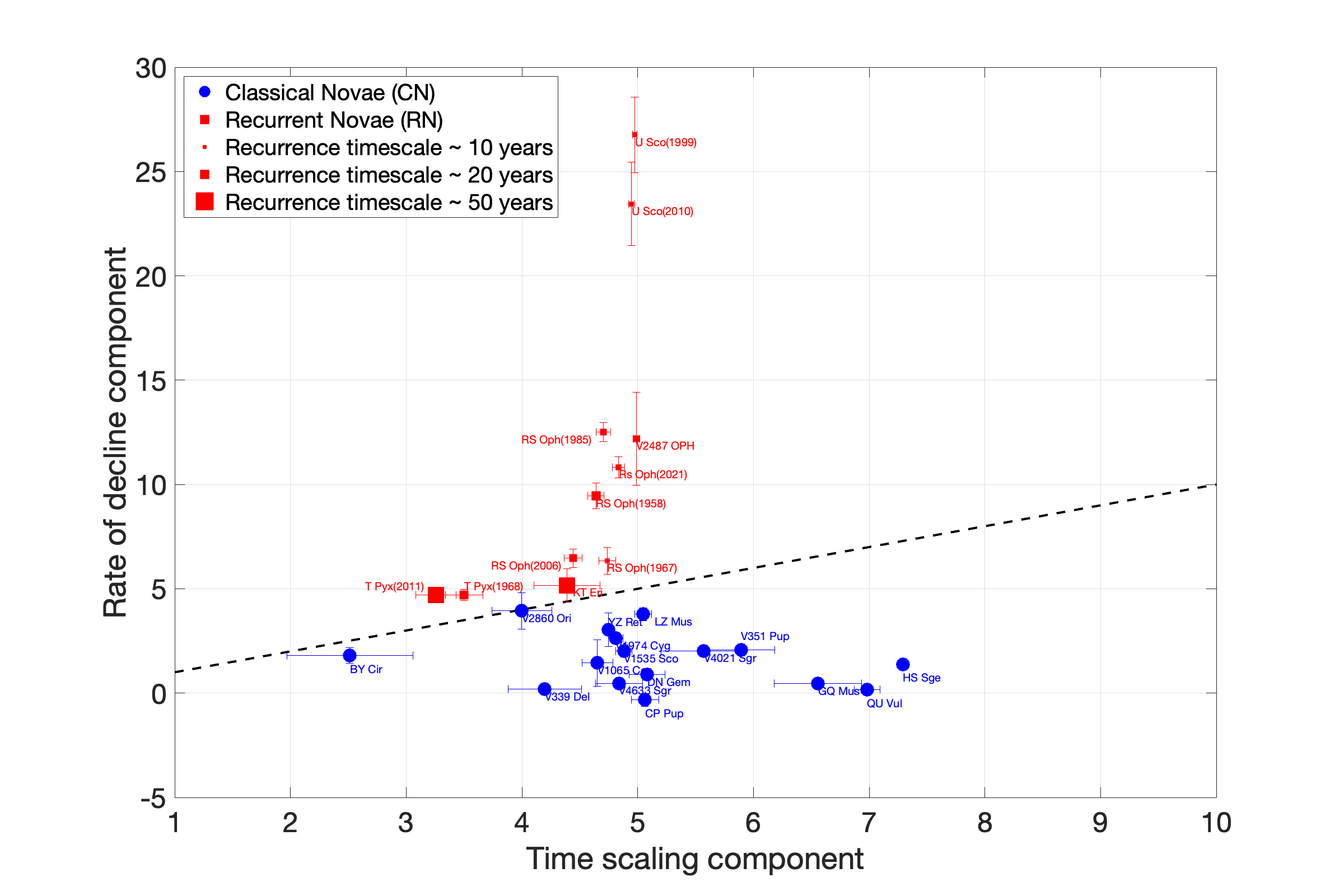}
  \caption{The position of RNe (red) and CNe (blue) in the parameter space of "rate of decline" vs. "time scaling" combination of light curve parameters. The dashed y = x line separates the parameter space into RNe and CNe regions. The point size for Recurrent Novae (RNe) are scaled to indicate their recurrence timescales (years between their latest and previous outburst). We see a clear trend of longer recurrence timescale closer to the separating line.}
\end{figure}

\subsection{KT Eri}

The first recorded eruption of KT Eri occurred in 2009, with no subsequent outbursts observed to date. Despite this single confirmed eruption, \citet{9} presented a compelling case for classifying KT Eri as a recurrent nova (RN), arguing that previous eruptions were likely missed due to observational gaps. Their analysis introduced eight observational diagnostics, which serve as proxies for the white dwarf mass and mass accretion rate—key parameters in nova classification.

More recently, \citet{2024MNRAS.529..224S} have observed a bright H$\alpha$-emitting shell centered on KT Eri, providing compelling evidence for the presence of a nova super-remnant (NSR) and reinforcing its classification as a RN. It has also been shown through hydrodynamic simulations by \citet{2024MNRAS.529..236H} that such a structure can naturally arise from frequent RN eruptions that efficiently clear out the surrounding interstellar medium, thereby offering a theoretical framework to interpret the observed remnant.

In our independent analysis of the light curve presented, we find that KT Eri consistently lies within the RN region of the parameter space, even when classification boundary is derived excluding KT Eri. Given the mounting observational and theoretical evidence supporting its RN classification, we adopt the final boundary obtained by including KT Eri as a recurrent nova. This result suggests that a simple linear classifier is sufficient to distinguish CNe from RNe within the current dataset. Nevertheless, more complex non-linear classifiers may be required as additional data become available in the future.

\section{Conclusions}

In this study, we present a phenomenological analysis of P-class optical light curves of Recurrent Novae (RNe) and Classical Novae (CNe), offering a practical framework for distinguishing between these two classes. A prior attempt, using just the light curve, to separate RNe from CNe using light curve properties was made by \citet{Duerback_1987}, based on decline timescales and outburst amplitudes. However, that method is limited by the requirement for quiescent photometry and does not always clearly distinguish RNe—for example, T Pyx falls in the CNe region of that parameter space. Our current approach appears to offer improved separability.

Given the inherent challenges in detecting multiple outbursts from the same system—due to undirected searches, weather conditions, and other observational constraints—confirmed discoveries of recurrent novae remain rare. Nevertheless, the method developed here shows potential in reliably distinguishing RNe from CNe based on light curve morphology alone. Notably, our analysis classifies KT Eri as a recurrent nova - from a completely independent study, consistent with the findings of \citet{Schaefer2022, 2024MNRAS.529..236H, 2024MNRAS.529..224S}. We also indicate the possibility of recurrence for V2860 Ori based on our analysis; deep observations searching for nova super-remnant, similar to that of KT Eri, may be useful to test this possibility.

While we employ a linear classification model in this work, yielding encouraging results, we recognize that the true separation between these nova types may be non-linear in nature. As more P-class nova light curves become available, future studies can build on this framework by utilizing the six extracted parameters in conjunction with more sophisticated machine learning classifiers. This will help refine the decision boundaries, improve classification accuracy, and deepen our physical understanding of nova evolution.

\section*{Acknowledgments}

We sincerely thank all the reviewers for their thoughtful comments and constructive suggestions, which helped improve the quality of this manuscript. We are also grateful to the numerous astronomers whose observations are available through the AAVSO portal, and to the AAVSO for maintaining this invaluable public database. We acknowledge Kautubh Kamal for initiating this project.

\newpage 

\appendix
\section{Linear fits for all the optical light curves}
\begin{figure}
  \centering
  \subfloat{\includegraphics[width=0.47\textwidth]{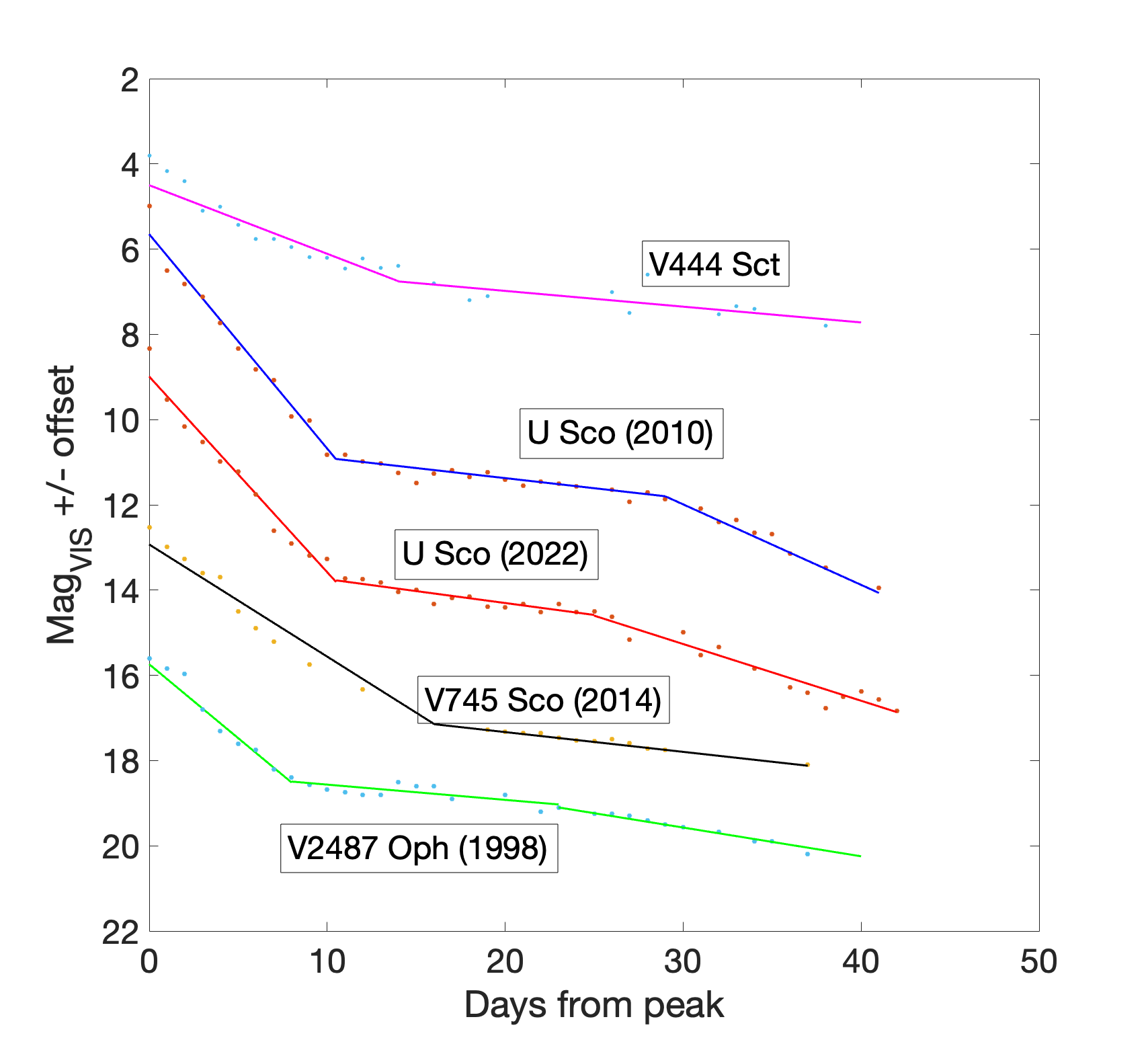}}\hfill
  \subfloat{\includegraphics[width=0.50\textwidth]{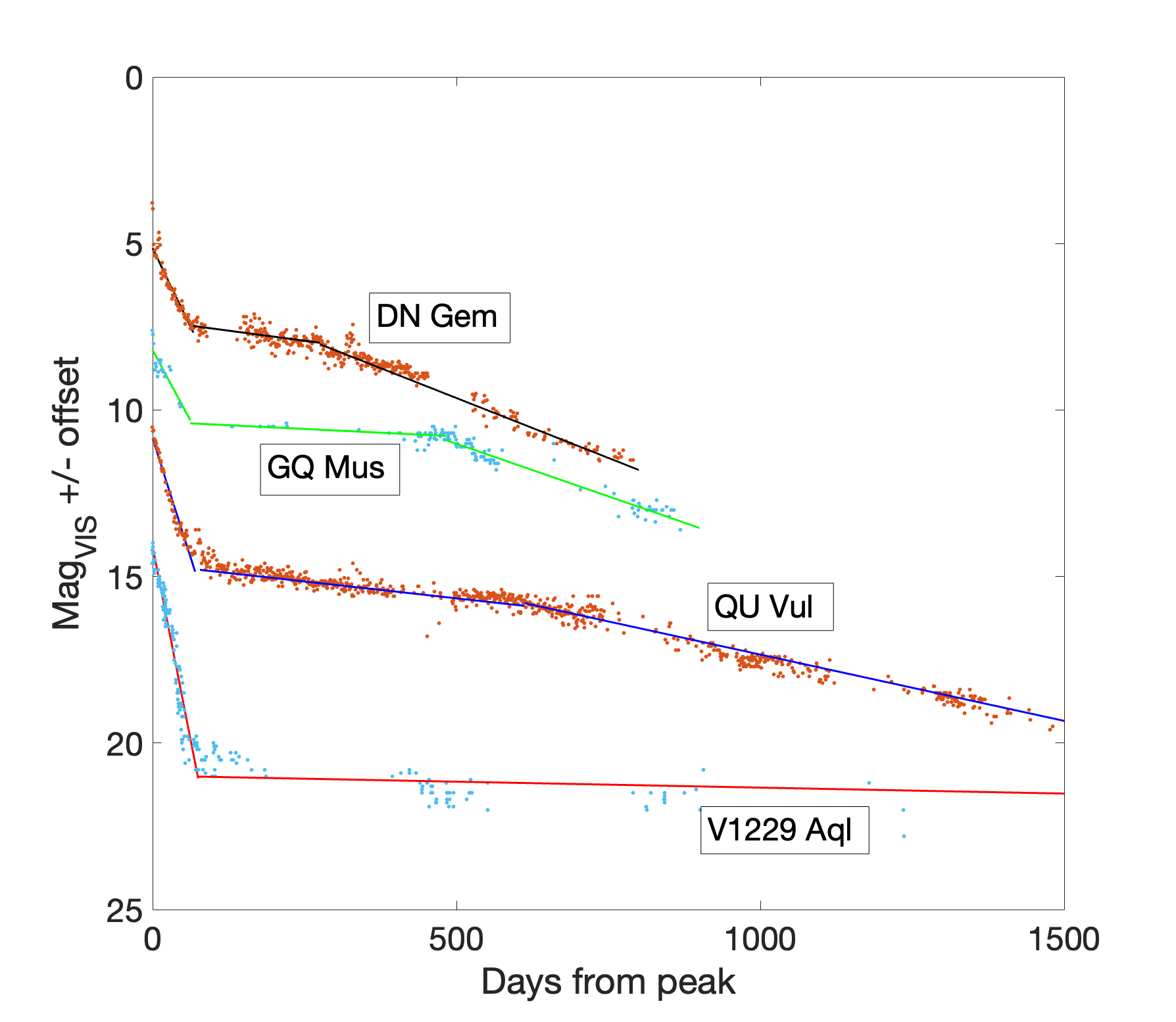}} \\
  \subfloat{\includegraphics[width=0.48\textwidth]{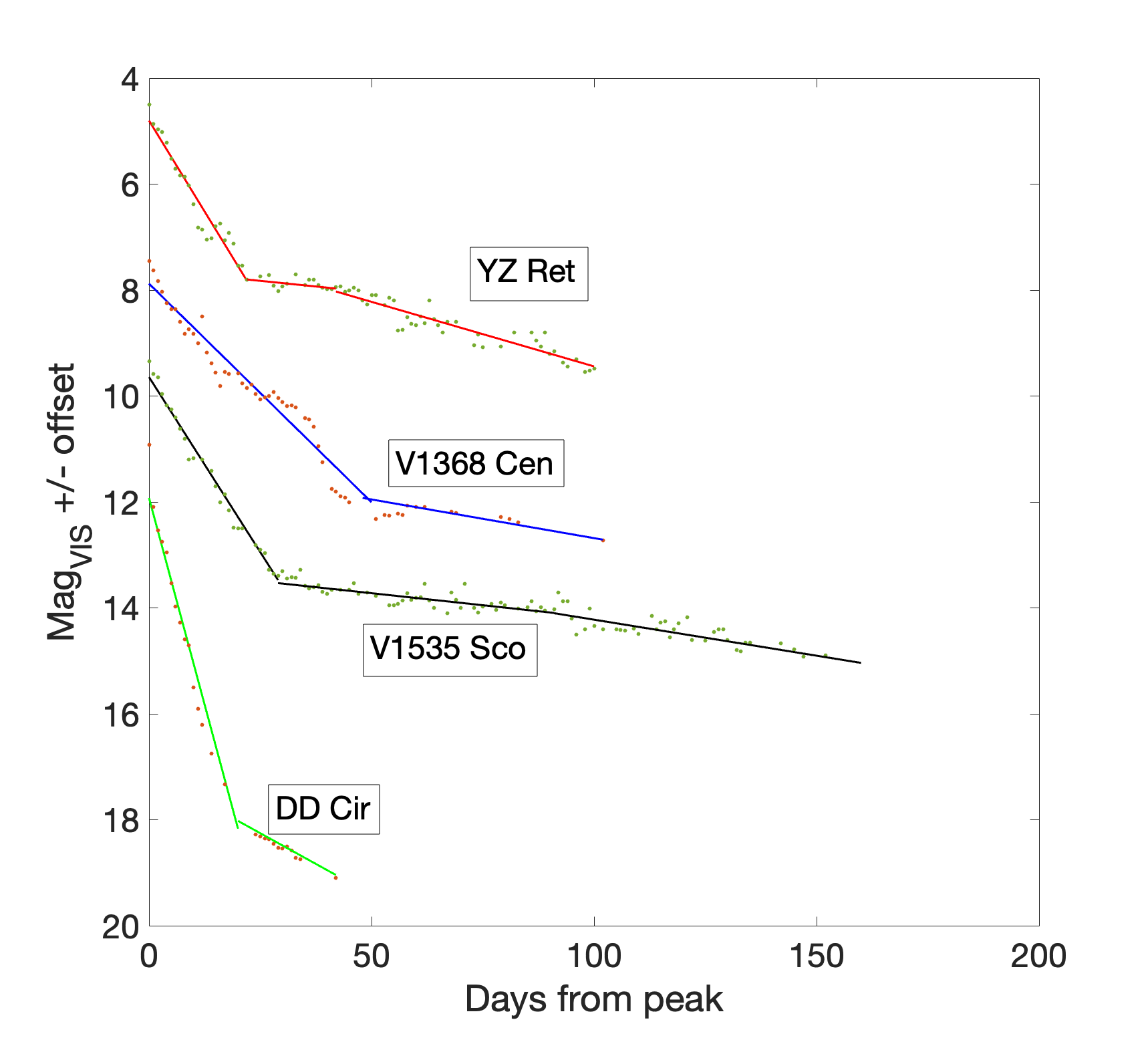}}\hfill
  \subfloat{\includegraphics[width=0.5\textwidth]{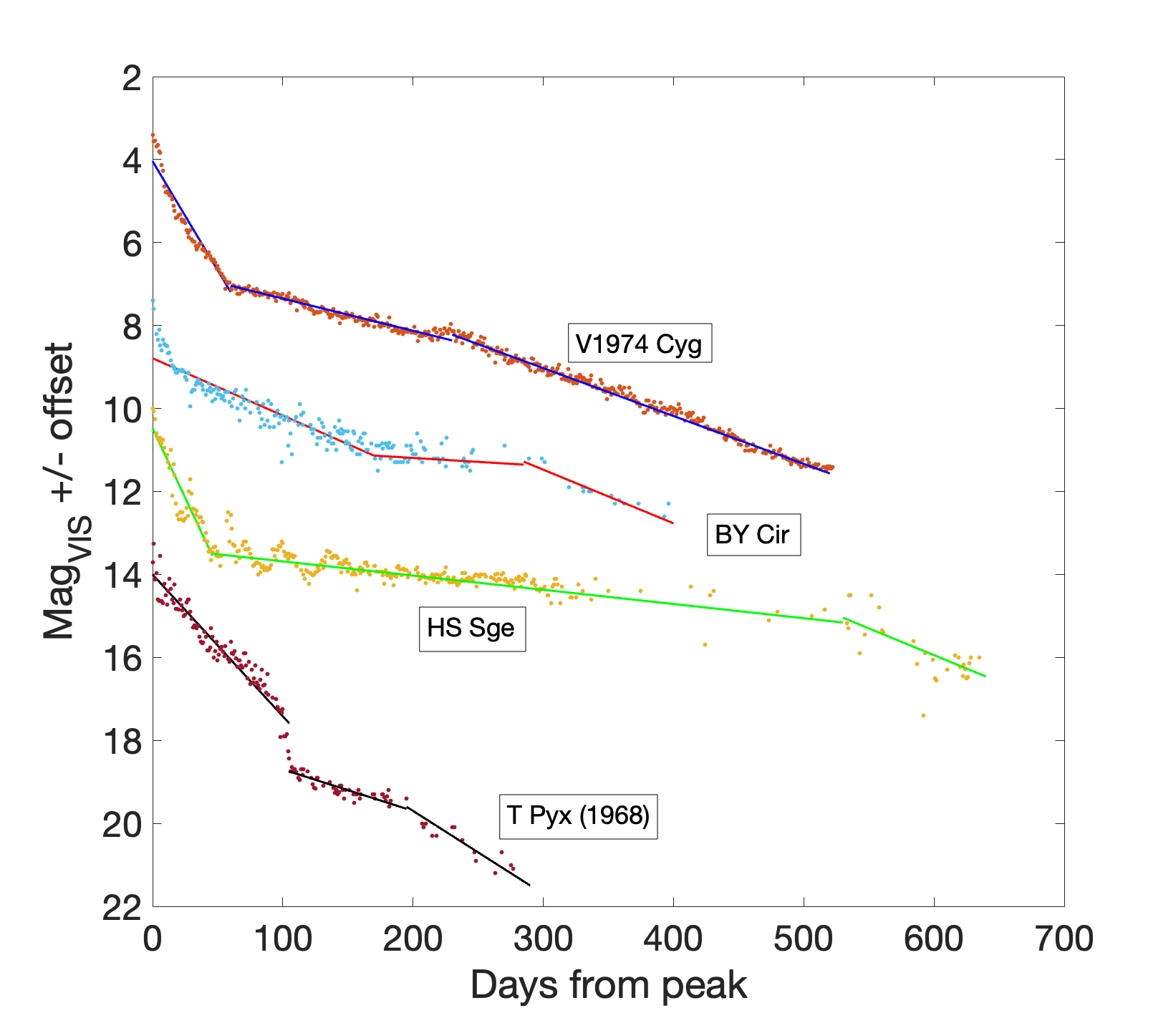}}
  \caption{}
\end{figure}

\begin{figure}
  \centering
  \subfloat{\includegraphics[width=0.47\textwidth]{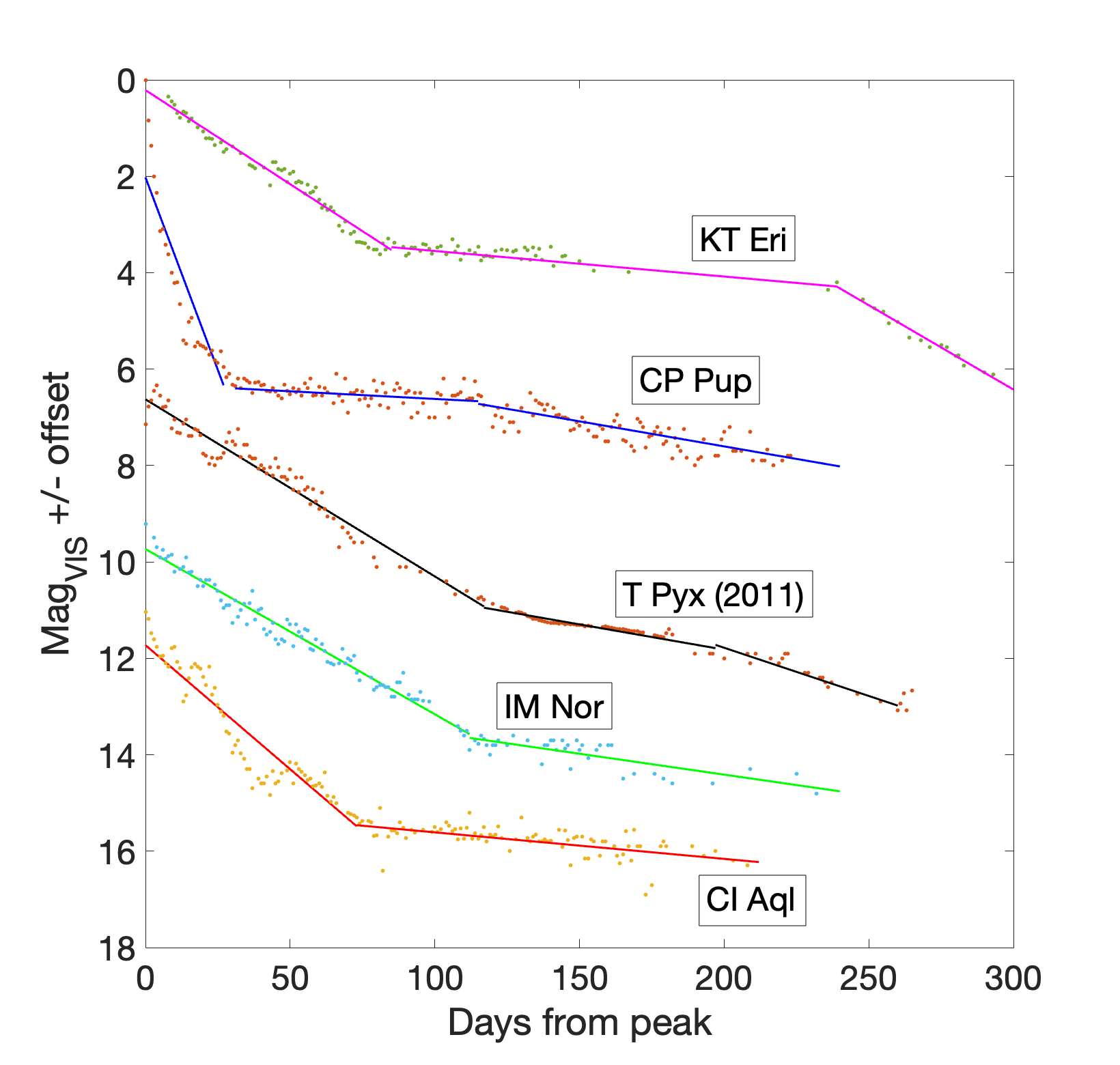}}\hfill
  \subfloat{\includegraphics[width=0.5\textwidth]{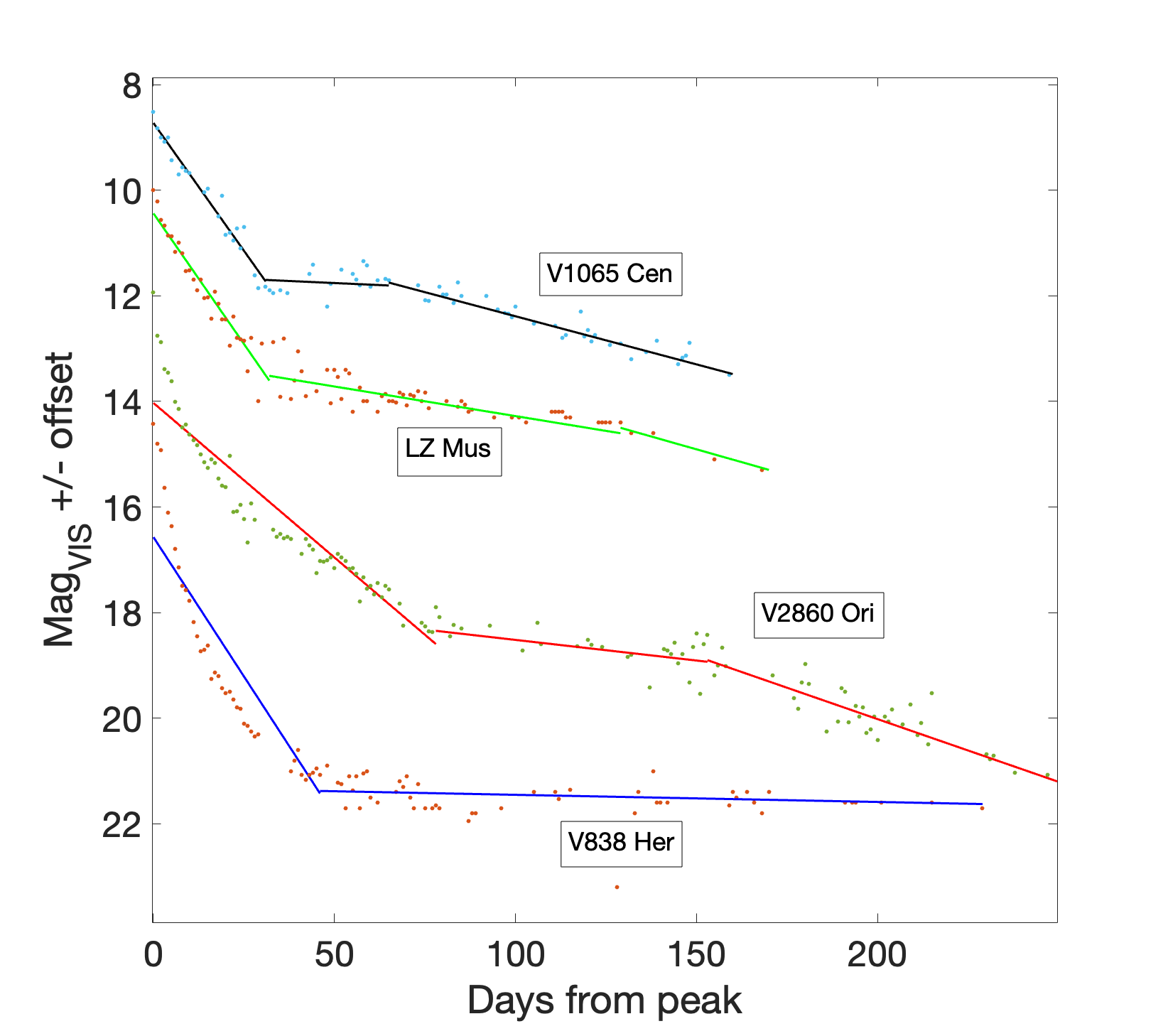}} \\
  \subfloat{\includegraphics[width=0.5\textwidth]{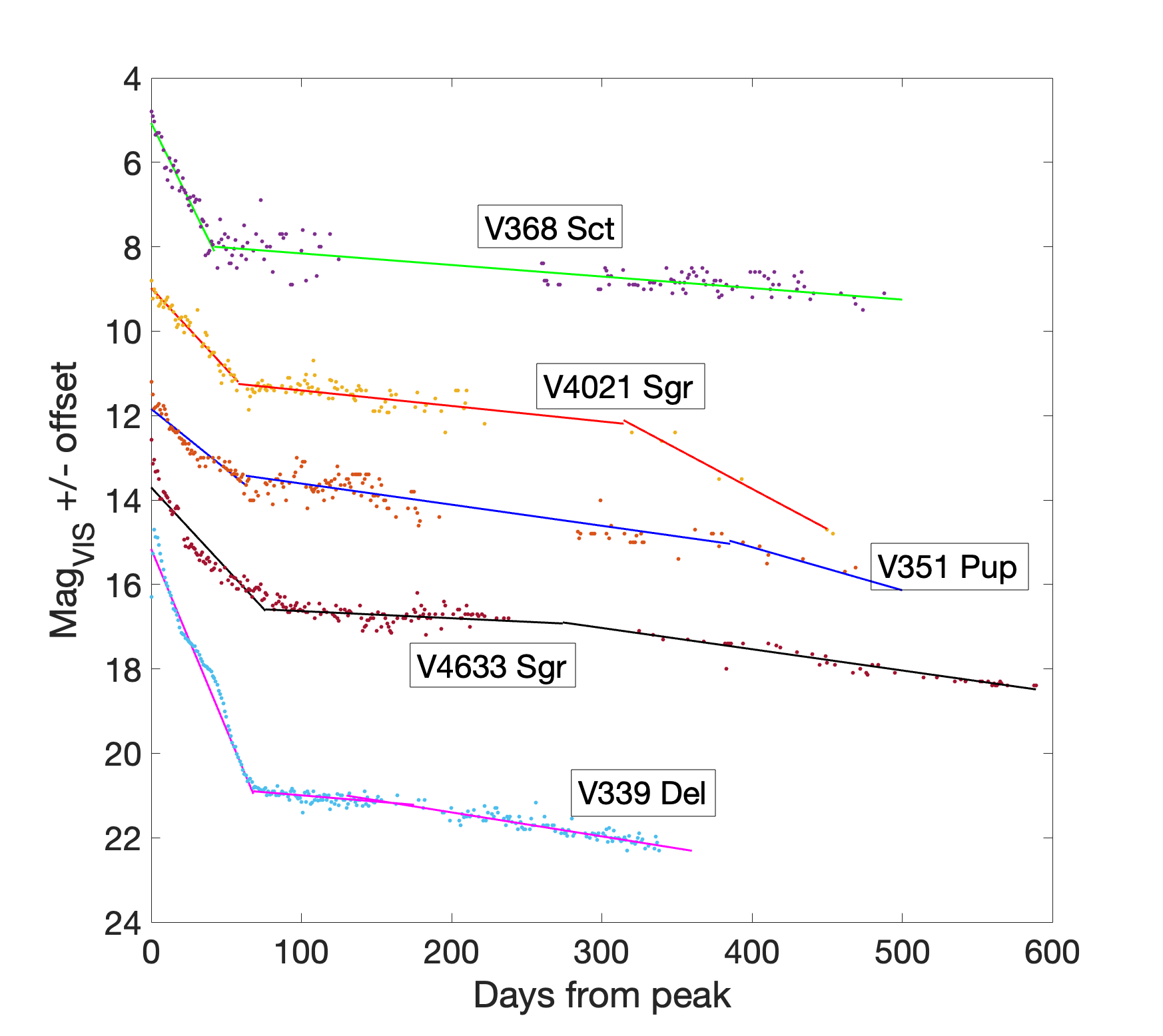}}\hfill
  \subfloat{\includegraphics[width=0.5\textwidth]{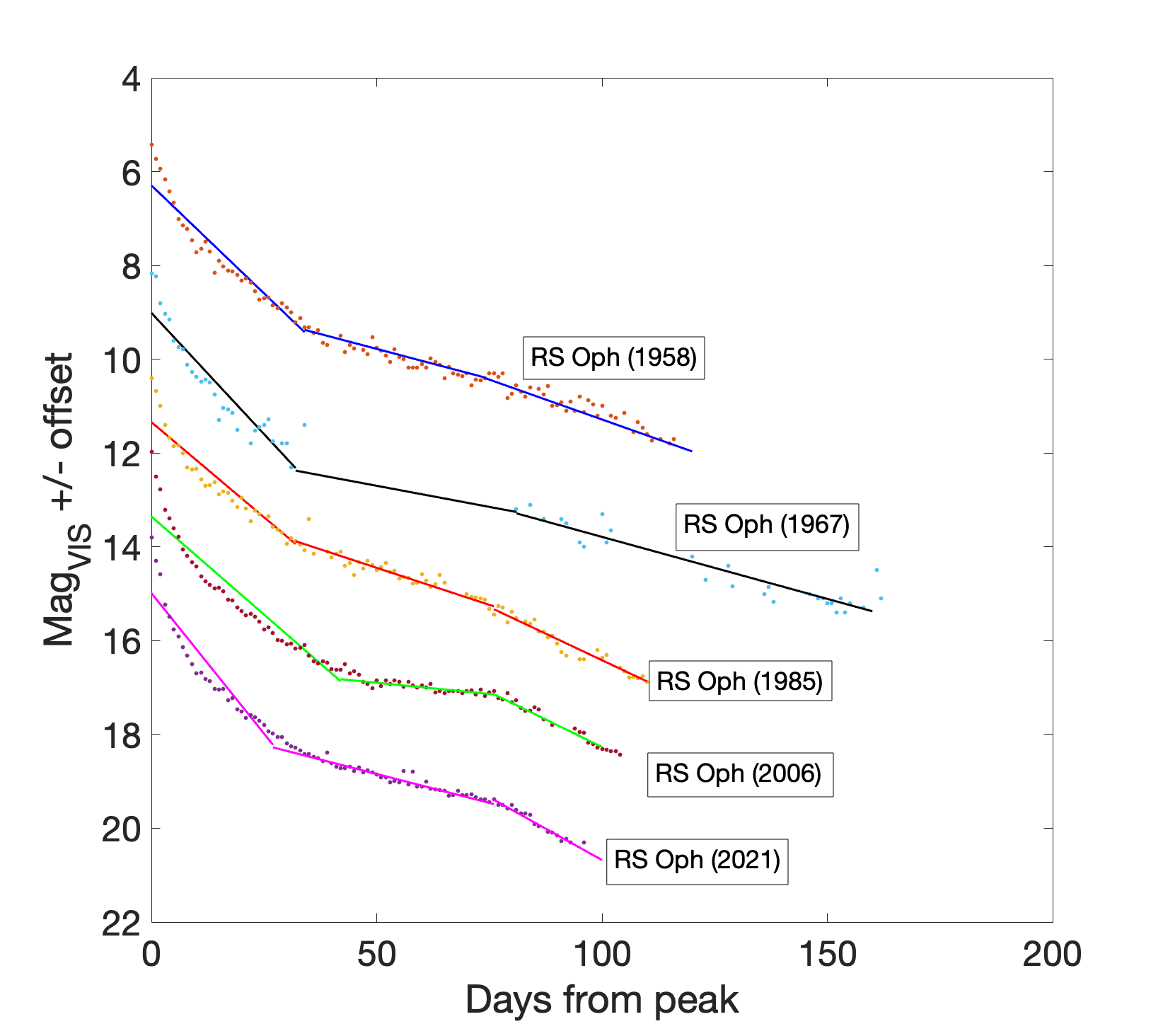}}
  \caption{}
\end{figure}

\bibliographystyle{jasr-model5-names}
\biboptions{authoryear}
\bibliography{jasr-template}

\end{document}